\documentstyle[12pt]{article}
\begin{document}
\thispagestyle{empty}
\noindent{\small{Theoretical and Mathematical Physics (1998) V.115. P.658} \\
\vspace{2cm}

\large
\vspace{1cm}
\begin{center}
\normalsize

{\Large \bf SPACETIME LOCALITY IN  $Sp(2)$ SYMMETRIC

\medskip

           \bf  LAGRANGIAN FORMALISM}

\vspace{3cm}

{\Large Shervgi S. Shahverdiyev$^{\dagger a}$ and I. V. Tyutin$^{\ddagger b}$ }

\vspace{1cm}

\vspace{3ex}

{$^a$ Physics Department,Lomonosov Moscow State University,}
\vspace{1ex}

{$^b$Lebedev Physical Institute,  Leninsky prospect 53, 117924 Moscow, Russia}\vspace{5ex}

\vspace{5ex}

\end{center}


\normalsize
\begin{quote}
The existence of a local solution to the $Sp(2)$  master equation for  gauge field theory
is proven in the framework of perturbation theory and under
standard assumptions on regularity of the action.
The arbitrariness of solutions to the $Sp(2)$  master equation
is described, provided that they are proper.
It is also shown that the effective action can be chosen  to be $Sp(2)$
and Lorentz invariant
(under the additional assumption that  the gauge
transformation generators are Lorentz tensors).
\end{quote}

\vfill
\noindent

$^\dagger$ E-mail address: shervgi@yahoo.com
\noindent

$^\ddagger$ E-mail address: tyutin@lpi.ru

\newpage

\section{Introduction}

It is well known that the  quantization methods for gauge theories in
Lagrangian and Hamiltonian formalisms can be generalized to include
ghost and antighost variables in a symmetric way \cite{05}--\cite{zy}.
In that case two nilpotent
(BRST and anti--BRST) charges appears. Ghost and antighost variables,
BRST and anti--BRST charges and the pair of equations for the effective action
form $Sp(2)$ doublets. In \cite{y}--\cite{d}, the formal proof of the
existence of a solution to the $Sp(2)$ master equation and description of the
arbitrariness of solutions were given.  In those papers, however, the
locality of the effective action was assumed as a hypothesis.  In
\cite{tsh},  the existence of a local solution to the $Sp(2)$ master equation
was demonstrated by using the Hamiltonian  $Sp(2)$ formalism in the case of
finite number of degrees of freedom (it was shown also that the Lagrangian and
Hamiltonian $Sp(2)$ formalisms are equivalent).

In the present paper, we shall prove
the existence of a local
solution to the $Sp(2)$ master equation in the field theory.
For the standard master equation of the BV formalism
the existence of a local
solution, the structure of renormalization, anomalies and related questions
 have been discussed in literature (see
\cite{p}, \cite{ht}, \cite{bbh} and references therein).
The method for solving the locality problem
is based on considering instead of equations
for functionals, equations for corresponding functions on spaces, which
have  the fields and their finite order derivatives  in an
arbitrary spacetime point as coordinates(jet--spaces \cite{Ss}).  We apply these
scheme to the $Sp(2)$ master equation.
In sec. 2 preliminary of the problem and assumptions on regularity of the
action are given.  In sec. 3 we give the proof of the existence of a local
solution to the $Sp(2)$ master equation.  In sec. 4 we discuss the
arbitrariness of solutions of the equations and show, under some restrictions
on the structure of the solution in the first order in the
ghost and gauge introducing fields, that the arbitrariness of
solutions is described by special transformations (gauge
transformations), which do not change the physical contents of the theory. It
is also shown that, under some additional assumptions on gauge generators,
there exist $Sp(2)$ symmetric and Lorentz invariant
solutions to the $Sp(2)$
master equation.

In what follows $\partial_\mu$ designates the total derivative with respect to
$x^\mu$, $
\partial_{\mu_1...\mu_k}\equiv\partial_{\mu_1}\cdots\partial_{\mu_k}$,
$D_{\mu_1...\mu_k}\equiv D_{\mu_1}\cdots D_{\mu_k}$.
By a local finite order differential operator (LFODO), we mean the
following
$$
M=\sum_{k=0}^n g^{\mu_1...\mu_k}\partial_{\mu_1...\mu_k},
$$
where $ g^{\mu_1...\mu_k}$ are functions of $x^\mu$, of the field variables and
their finite order derivatives, $n$ is
a finite number. Transposed LFODO $M^{Tij}$ relative to
$M$ is defined as
$$
(M^{Tij}\varphi_j)\psi_i-\varphi_iM^{ij}\psi_j=\partial_\mu j^\mu,\quad
(M^{-1})^T=(M^T)^{-1},\quad
(M^{T})^{T}=(-1)^{\varepsilon_i+\varepsilon_j}M^{ij},
$$
$$
\quad
\varepsilon(\varphi^i)=\varepsilon_i,\quad \varepsilon(M^{ij})=
\varepsilon_i+\varepsilon_j
$$
with arbitrary functions $\varphi_i$, $\psi_i$ and local functions $j^\mu$.
All considerations are performed in the framework of perturbation theory,
i.e., in terms of formal power series in field variables and
their derivatives. Throughout this article all functions can depend explicitly
on $x^\mu$, unless otherwise is stated.

\section{Preliminary of the problem}

In this  section we shall discuss the formulation of the problem and
assumptions under which it will be solved. The full set of variables
$\Gamma^\Sigma$ is divided into the groups  $ \Phi^A, \Phi_{Aa}^*,
{\bar\Phi}_A,$ $ a, b, c=1, 2$. The variables $\Phi^A $ are $\Phi^A
=(\varphi^i,C^{\alpha b},B^\alpha)$, where $\varphi^i$ are  variables  of
the classical theory, while $ C^{\alpha b}$ and $B^\alpha $ are respectively
ghost, antighost and gauge introducing fields. All variables are ascribed by
the Grassman parity $
\varepsilon(\Phi^A)=\varepsilon(\bar{\Phi}_A)=\varepsilon_A$, $
\varepsilon(\Phi^*_{Aa})=\varepsilon_{A}+1$,
$\varepsilon(B^\alpha)=
\varepsilon_\alpha$, $\varepsilon(C^{\alpha
b})=\varepsilon_\alpha+1
$
and the new ghost number ngh:
$\hbox{ngh}(\varphi)=0$, $\hbox{ngh}(C^b)=1$, $\hbox{ngh}(B)=2$,
$\hbox{ngh}(\bar{\Phi})=-2-\hbox{ngh}(\Phi)$,
$\hbox{ngh}(\Phi^*_a)=-1-\hbox{ngh}(\Phi)$,
$ \hbox{ngh(FM)}=\hbox{ngh(F)}+\hbox{ngh(M)}$.
Moreover, the fields $\varphi^i, B^\alpha, \bar{\varphi_i}, \bar{B_\alpha}$
form $Sp(2)$ singlets. $ C^{\alpha b} $ and
 $\varphi^*_{ia} $, $ B^*_{\alpha a} $, $ \bar{C}_{\alpha a}$ form respectively $Sp(2)$  doublets and antidoublets, $C^*_{\alpha b|a}
$
transforms as a product of $Sp(2)$ antidoublets.
In the $Sp(2)$ formalism the effective action
 $ S(\Phi, \Phi_a^*, {\bar\Phi}) $
satisfies  the $Sp(2)$ master equation

\begin{equation}\label{1}
\frac{1}{2}(S,
S)^a+
\int dx\varepsilon^{ab}\Phi_{Ab}^*
    \frac{\delta}{\delta{\bar\Phi}_{A}}S=0,
\quad
\varepsilon^{ab}=-\varepsilon^{ba},\quad
\varepsilon_{ab}\varepsilon^{bc}=\delta^c_a,\quad
\varepsilon_{12}=\varepsilon^{21}=1,
\end{equation}
and the boundary condition
$$
\left.S\right|_{\Phi_{a}^*={\bar\Phi}=0}=\cal S(\varphi).
$$
In (\ref{1}), $(, )^a$ denotes the doublet of antibrackets
$$
(F,G)^a\equiv\int dx\left( F{\overleftarrow{\delta}\over\delta\Phi^A}
{\overrightarrow{\delta}\over\delta\Phi^*_{Aa}}G-
F{\overleftarrow{\delta}\over\delta\Phi^*_{Aa}}
{\overrightarrow{\delta}\over\delta\Phi^A}G\right),\nonumber
 $$
where
${\cal S}(\varphi)$ is the original classical action, which has a gauge
symmetry:  $ R^i_\alpha \delta {\cal S}/\delta \varphi_i=0.  $ It is assumed
that the effective action conserves ngh, has zero Grassman parity
 and forms $Sp(2)$ singlet.  Equation (\ref{1}) will be solved in the
 framework of perturbation theory in fields
 $ C^{\alpha b}$ and $B^\alpha$:
$$ S={\cal S}+\sum_{k=1}S_k, \quad S_k \sim
C^l B^{m}, \quad l+m=k.
$$
Let us assume that the terms $ S_k,$ are proved to exist for $k=1,..., n
$, that is, equation (\ref{1}) is satisfied in orders $C^l B^{k-l}$, $k\le n$.
In the $(n+1)$th  order in fields $ C^{\alpha b}$ and $B^\alpha$,
we obtain equation for $S_{n+1}$:
\begin{equation}\label{2}
W^aS_{n+1}=F_{n+1}^a, \quad F_{n+1}^a=-\frac{1}{2}(S_{[n]},
S_{[n]})^a_{n+1}, \quad
S_{[n]}={\cal S}+\sum_{k=1}^n S_k,
\end{equation}
where (see next Sec.):
$$ W^a=\int
dx\left((-1)^{\varepsilon_i}L_i\frac{\delta}{\delta \varphi^*_{ia}} -
(-1)^{\varepsilon_\alpha}R^i_\alpha \varphi^*_{ib}\frac{\delta}{\delta
C^*_{\alpha b|a}} +
((-1)^{\varepsilon_i}R^i_\alpha\bar{\varphi_i}+\varepsilon^{cb}C^*_{\alpha
b|c}) \frac{\delta}{\delta B^*_{\alpha a}}
\right.
$$
$$
\left.
-(-1)^{\varepsilon_\alpha}
\varepsilon^{ab}B^\alpha\frac{\delta}{\delta C^{\alpha b}}
+
\varepsilon^{ab}\Phi^*_{Ab}\frac{\delta}{\delta \bar{\Phi}_A}
\right),
$$
$$
L_i(x)\equiv
\delta {\cal S}/\delta
\varphi_i(x).
$$
The operators  $W^a$ are nilpotent, i.e.,
$$
W^aW^b+W^bW^a=0.
$$
$W^a$ and $F^a_{n+1}$ form $Sp(2)$ doublets.
The Jacobi identity for doublet of antibrackets or explicit form of
$F^a_{n+1}$ implies
the consistency condition
\begin{equation}\label{3}
W^aF^b_{n+1}+W^bF^a_{n+1}=0,
\end{equation} for equation (\ref{2}).  One can see that
\begin{equation}\label{4} F^a_{n+1}=W^aY_{n+1}
\end{equation} is a solution
to equation (\ref{3}).  If this is the  general solution we may
choose  $Y_{n+1}$ as $S_{n+1}$.  We shall show, that under some
assumptions formulated below the general solution to equation
(\ref{3}) indeed has the form (\ref{4}).  The main difficulty  in
solving equation (\ref{4}) is to obtain a solution belonging to the class
of local functionals, i. e., the functionals $S_k$
 should have the
form $S_k=\int dx s_k(\Gamma^\Sigma, \partial_\mu\Gamma^\Sigma,...)$.
Consequently, $F_{n+1}^a$ will be  $F^a_{n+1}=\int dx
f^a_{n+1}(\Gamma^\Sigma, \partial_\mu\Gamma^\Sigma,...)$
\footnote{ The functions $s_k$ and $f^a_{n+1}$ are defined up to a total derivative. }.  The operators $W^a$ on
arbitrary local functions are:  \begin{equation} \label{5} \begin{array}{l}
W^aZ=\displaystyle\int dx w^a z(\Gamma^\Sigma, \partial_\mu
\Gamma^\Sigma, ...), \\
w^a=
\displaystyle\sum_{k=0}(-1)^{\varepsilon_i}D_{\mu_1...\mu_k}L_i\frac{\partial}{\partial
(\partial_{\mu_1...\mu_k}\varphi^*_{ia})}
-
\displaystyle\sum_{k=0}(-1)^{\varepsilon_\alpha}D_{\mu_1...\mu_k}(R^i_\alpha\varphi^*_{ib})\frac
{\partial}{\partial
(\partial_{\mu_1...\mu_k}C^*_{\alpha b|a})}
+          \\
\displaystyle\sum_{k=0}D_{\mu_1...\mu_k}((-1)^{\varepsilon_i}
R^i_\alpha\bar{\varphi_i}+\varepsilon^{cb}C^*_{\alpha b|c})
\frac{\partial}{\partial
(\partial_{\mu_1...\mu_k}B^*_{\alpha a})}
-
(-1)^{\varepsilon_\alpha}
\varepsilon^{ab}
\displaystyle\sum_{k=0}\partial_{\mu_1...\mu_k}B^\alpha\frac
{\partial}{\partial
(\partial_{\mu_1...\mu_k}C^{\alpha b})}
+          \\
\varepsilon^{ab}
\displaystyle\sum_{k=0}\partial_{\mu_1...\mu_k}\varphi^*_{ib}\frac{\partial}{\partial
(\partial_{\mu_1...\mu_k}\bar{\varphi}_i)}
+\varepsilon^{ab}
 \displaystyle\sum_{k=0}\partial_{\mu_1...\mu_k}B^*_{\alpha b}
\frac{\partial}{\partial
(\partial_{\mu_1...\mu_k}\bar{B}_{\alpha})}
+\varepsilon^{ab}
\displaystyle\sum_{k=0}\partial_{\mu_1...\mu_k}C^*_{\alpha c|b}
\frac{\partial}{\partial
(\partial_{\mu_1...\mu_k}\bar{C}_{\alpha c})}, \\
D_\mu\equiv\displaystyle\frac{\partial}{\partial x^\mu}+
\displaystyle\sum_{k=0}\partial_{\mu\mu_{1}...\mu_{k}}\Gamma^\Sigma
\displaystyle{\partial\over\partial(\partial_{\mu_{1}...\mu_{k}}\Gamma^\Sigma)},
\end{array}
\end{equation}
all quantities in (\ref{5}) are taken in an arbitrary fixed spacetime point
$x^\mu$.  The operators $w^a$ and $D_\mu$ commute:  $$
w^aw^b+w^bw^a=[D_\mu,w^a]=[D_\mu, D_\nu]=0.
$$
Equation (\ref{3}) in terms of nonintegrated densities is
\begin{equation}\label{6}
w^af^b_{n+1}+w^bf^a_{n+1}=D_\mu j^{\mu ab}_{n+1},
\end{equation}
where
$
j^{\mu ab}_{n+1}=j^{\mu ab}_{n+1}(\Gamma^\Sigma, \partial_
\mu \Gamma^\Sigma, ...)
$
are functions of $\Gamma^\Sigma$
and their derivatives up to finite order.
To solve equation (\ref{6}), it is helpful to introduce operators $\gamma_a$,
which contain derivatives
$
{\partial/\partial L_i}$, $\partial/\partial
(R^i_\alpha \varphi^*_{ia})
$.
To make the introducing of such operators possible we make the
standard
assumptions on the structure of the action $\cal S$ and
of the gauge generators $R^i_\alpha$, which we will call
the regularity assumptions of the theory.

1. Gauge transformation generators are LFODO's:
\begin{equation} \label{7}
  \begin{array}{rcl}
R^i_\alpha=r^i_\alpha+ r^{i\mu}_\alpha \partial_\mu
+...+r^{i\mu_1\mu_2...\mu_t}_\alpha \partial_{\mu_1\mu_2...\mu_t}, \end{array}
\end{equation}
$r^{i\mu_1\mu_2...\mu_k}_\alpha$, $k=0,...,t$,
are functions of the classical fields
and their derivatives up to finite order,
$\varepsilon(R^i_\alpha)=\varepsilon(r^{i\mu_1\mu_2...\mu_t}_\alpha)=
\varepsilon_i+\varepsilon_\alpha.$

2. Generators $R^i_\alpha$ form an irreducible set in the sense, that any
LFODO $\Lambda^i$, satisfying the equation $$ \Lambda^iL_i=0, $$ has the form
$$ \Lambda^i=m^\alpha R^i_\alpha+\hat{M}^{ij}L_j, \quad
\hat{M}^{ij}A_j=\sum_{k, l}
m^{i\mu_1...\mu_k|j\nu_1...\nu_l}
\partial_{
\nu_1...\nu_l}A_j\partial_{\mu_1...\mu_k},
$$
$$
m^{i\mu_1...\mu_k|j\nu_1...\nu_l}=-(-1)^{\varepsilon_i\varepsilon_j}
m^{j\nu_1...\nu_l|i\mu_1...\mu_k}
$$
where $m^\alpha$ are LFODO's, $m^{i\mu_1...\mu_k|j\nu_1...\nu_l}$ are
functions of $\Gamma^\Sigma$
and of their derivatives up to a finite order. Furthermore, equations
$R^{Ti}_\alpha n^\alpha=0$ have the only solution $n^\alpha=0$.

3. Let us consider the space $J_l(\varphi)$ (jet space) with coordinates
$ \varphi_l\equiv(\varphi^i, \partial_\mu
\varphi^i, \partial_{\mu_1}....\partial_{\mu_l}\varphi^i)$ .
Given set of functions $ L_Q\equiv(L_i, D_\mu L_i, ..., D_{\mu_1...
\mu_{\sigma_l}}L_i)$,
$Q=(i, i\nu_1,..., i\nu_1...\nu_{\sigma_{l}})$,
such that the order of
the highest derivatives of fields $\varphi^i$
occurring in this set is equal to $l$.
There exist constraints among the functions $L_Q$
\footnote{In (\ref{8}), $ R^i_\alpha $ are given by
(\ref{7}), with
$\partial_\mu\to D_\mu $.}:
\begin{equation} \label{8}
  \begin{array}{rcl}
R^i_\alpha L_i=0,\quad D_\mu\left(R^i_\alpha L_i\right)
=0, ..., \quad
D_{\mu_1...  \mu_{s_l}}\left(R^i_\alpha L_i\right)=0,
\end{array}
\end{equation}
where $s_l$ is chosen from the condition that
the highest derivatives of fields $\varphi^i$
occurring in this set be equal to $l$
\footnote{In general, $s_l$ and $t$ depend on $i$ and $\alpha$ respectively.
In the present paper, we omit these nonessential questions.}.
Let us rewrite the constraints in the form
$$ \tilde{R}^Q_D L_Q=\tilde{R}^I_D L_I+\tilde{R}^{D^\prime}_D
L_{D^\prime}=0,
$$
where
$D=(\alpha,\alpha\nu_1,...,\alpha\nu_{1}...\nu_{s_l})$,
$\tilde{R}^I_D $ and
$\tilde{R}^{D^\prime}_D$ are matrices, depending on $\varphi_l $.
We assume that for any $l$ the set $L_Q$ can be divided into sets of
independent functions $L_I $ ($[L_I]$ designates the number of $L_I$) $$
\left.rank\left(\frac{\partial
L_I}{\partial \varphi_l}\right)\right|_{L_Q=0}=[L_I]
$$
 and dependent ones
$ L_{D^\prime}$, $\tilde{R}^{D^\prime}_D$ being a nonsingular matrix.

In sections  3 and 4 we will show that under the regularity
 assumptions of the theory
the $Sp(2)$ master equation has a local solution.

\section{The existence of a local solution}

One can check that the functional
\begin{equation}\label{9}
S_1=\int dx\left(C^{\alpha a}R^i_\alpha
\varphi^*_{ia}+\varepsilon^{ab}
C^*_{\alpha b|a}B^\alpha+B^\alpha
R^i_\alpha\bar{\varphi_i}(-1)^{\varepsilon_i+\varepsilon_\alpha}\right)
\end{equation}
is a solution to the
 $Sp(2)$ master equation
in the first order in $C^{\alpha b}, B^\alpha$.
Let us investigate the structure of the general solution to equation (\ref{6}).
All functions $s_k$, $k\le n$
contain derivatives of $\Gamma^\Sigma$
up to a finite order.
Consequently, the functions $f^a_{n+1}$ also contain finite order derivatives
of $\Gamma^\Sigma$.
We choose the jet space $
J_{(l)}(\Gamma^\Sigma)=J_l(\varphi)\bigotimes J_{l_C}(C)\bigotimes...  $, $
l_{\varphi^*_a}=l_{\bar{\varphi}}=\sigma_l,
l_{B^*_a}=l_{C^*_{b|a}} =l_{\bar{B}}=l_{\bar{C_a}}=s_l,
l_{C^a}=l_B $ in such a way that  $f^a_{n+1}$,
$w^af^b_{n+1}$, $D_\mu j^{\mu ab}$ be defined on it.
Note, that if we restrict ourselves to the operators
$w^a$ and $D_\mu$ as only acting on functions defined on
$J_{(l)}(\Gamma^\Sigma)$, then all sums in expressions (\ref{5}) and
(\ref{6})
are finite.
Of course, jet spaces are defined ambiguously: if a set $(l)$
is admissible, then any other set $(l^\prime), l^\prime\ge l,
l^\prime_C\ge l_C, ... $ is admissible, too.

The operators $w^a$ acting on $J_{(l)}(\Gamma^\Sigma)$ are
$$
w^a=
(-1)^{\varepsilon_I}L_I\frac{\partial}{\partial
\varphi^*_{Ia}}
+
(-1)^{\varepsilon_{D^\prime}}L_{D^\prime}\frac{\partial}{\partial
\varphi^*_{D^\prime a}}
-
(-1)^{\varepsilon_D}\tilde{R}^Q_D\varphi^*_{Qb}\frac{\partial}{\partial
C^*_{Db|a}}
+
((-1)^{\varepsilon_Q}\tilde{R}^Q_D\bar{\varphi}_{Q}+\varepsilon^{bc}C^*_{Dc|b})
\frac{\partial}{\partial
B^*_{Da}}
-
$$
$$(-1)^{\varepsilon_\alpha}
\varepsilon^{ab}
\sum_{k=0}\partial_{\mu_1...\mu_k}B^\alpha\frac
{\partial}{\partial
(\partial_{\mu_1...\mu_k}C^{\alpha b})}
 +\varepsilon^{ab}
\varphi^*_{Qb}\frac{\partial}{\partial
\bar{\varphi}_Q}
+
\varepsilon^{ab}
 B^*_{Db}
\frac{\partial}{\partial
\bar{B}_{D}}
+\varepsilon^{ab}
C^*_{Dc|b}
\frac{\partial}{\partial
\bar{C}_{Dc}}.
$$
After the change of variables
$$
(\varphi^{*}_{Ia}, \varphi^*_{D^\prime
a})\to(\varphi^{*}_{Ia}, \varphi^{\prime*}_{Da}=
\tilde{R}^Q_D\varphi^*_{Qa}),\quad
(\bar{\varphi}_I,
\bar{\varphi}_{D^\prime})\to
(\bar{\varphi}_I,
\bar{\varphi}_D^\prime=(-1)^{\varepsilon_Q}\tilde{R}^Q_D\bar{\varphi}_{Q}+
\varepsilon^{bc}C^*_{Dc|b}),
$$
with the rest of the variables left unchanged, the operators $w^a$
take the form (the primer is omitted)
$$
w^a=(-1)^{\varepsilon_I} L_I\frac{\partial}{\partial \varphi^*_{Ia}}
+\varepsilon^{ab}
\varphi^*_{Ib}\frac{\partial}{\partial
\bar{\varphi}_I}
-
(-1)^{\varepsilon_D}\varphi^*_{Db}\frac{\partial}{\partial
C^*_{Db|a}}
+
\varepsilon^{ab}
C^*_{Dc|b}
\frac{\partial}{\partial
\bar{C}_{Dc}}
+\bar{\varphi}_{D}\frac{\partial}{\partial
B^*_{Da}}
+
$$
$$
\varepsilon^{ab}
 B^*_{Db}
\frac{\partial}{\partial
\bar{B}_{D}}
-(-1)^{\varepsilon_\alpha}
\varepsilon^{ab}
\sum_{k=0}\partial_{\mu_1...\mu_k}B^\alpha\frac
{\partial}{\partial
(\partial_{\mu_1...\mu_k}C^{\alpha b})}.
$$
Next, we introduce the operators
$$
\gamma_a=
(-1)^{\varepsilon_I}\varphi^*_{Ia}\frac{\partial}{\partial
L_I}
-
\varepsilon_{ab}
\bar{\varphi}_I\frac{\partial}{\partial\varphi^*_{Ib}
}
-
(-1)^{\varepsilon_D}C^*_{Db|a}\frac{\partial}{\partial
\varphi^*_{Db}}
-\varepsilon_{ab}
\bar{C}_{Dc}
\frac{\partial}{\partial
C^*_{Dc|b}}
+
$$
$$
B^*_{Da}\frac{\partial}{\partial
\bar{\varphi}_{D}}
-
\varepsilon_{ab}
\bar{B}_{D}\frac{\partial}{\partial
B^*_{Db}}.
$$
Simple calculations give the following relations
$$
\gamma_a\gamma_b+\gamma_b\gamma_a=0,\quad
w^a\gamma_b+\gamma_bw^a=\delta^a_bN,\quad
w^aN=Nw^a,\quad\gamma_aN=N\gamma_a,
$$
$$
N=
L_I\frac{\partial}{\partial
L_I}
+
\varphi^*_{Qb}\frac{\partial}{\partial
\varphi^*_{Qb}}
+
\bar{\varphi}_{Q}\frac{\partial}{\partial
\bar{\varphi}_{Q}}+
B^*_{Db}\frac{\partial}{\partial
B^*_{Db}}+
C^*_{Db|a}
\frac{\partial}{\partial
C^*_{Db|a}}
+
\bar{B}_{D}
\frac{\partial}{\partial
\bar{B}_{D}}+
\bar{C}_{Dc}
\frac{\partial}{\partial
\bar{C}_{Dc}}.
$$
$D_\lambda$ can be written as
$D_\lambda=\bar{D}_\lambda+\tilde{D}_\lambda$,
$$
\bar{D}_\lambda=
 \sum_{k=1}\partial_{\mu_1...\mu_k}C^{\alpha a}
\frac{\partial}{\partial
(\partial_{\mu_1...\mu_{k-1}}C^{\alpha a})}+
 \sum_{k=1}\partial_{\mu_1...\mu_k}B^{\alpha}
\frac{\partial}{\partial
(\partial_{\mu_1...\mu_{k-1}}B^{\alpha})},
$$
$$
[w^a, \bar{D}_\lambda]=
[\gamma_a, \bar{D}_\lambda]=
[N, \bar{D}_\lambda]=0,\quad
[w^a, D_\lambda]=
[N, D_\lambda]=0,\quad
[\gamma_a, D_\lambda]\neq 0.
$$
Consider equation (\ref{6}).  To solve it we apply the method used in \cite{H}
for the standard formalism. Let us expand $f^a_{n+1}$ and $j^{\mu
ab}_{n+1}$ according to the total number of derivatives of the fields
$ C^{\alpha a}$ and $
B^{\alpha}$:
$$ f^a_{n+1}=\sum_{k=0}^{p}f^{(k)a}_{n+1},\quad j^{\lambda
ab}_{n+1}=\sum_{k=0}^{m}j^{(k)\lambda ab}_{n+1},\quad
Df^{(k)a}_{n+1}=kf^{(k)a}_{n+1},\quad
Dj^{(k)\lambda ab}_{n+1}=kj^{(k)\lambda ab}_{n+1},
$$
$$
D=\sum_{k=0}k\partial_{\mu_1...\mu_k}
C^{\alpha
a}\frac{\partial}{\partial
(\partial_{\mu_1...\mu_k}C^{\alpha a})}
+
\sum_{k=0}k\partial_{\mu_1...\mu_k}B^\alpha\frac
{\partial}{\partial
(\partial_{\mu_1...\mu_k}B^{\alpha })},
$$
where $p$ and $m$ are finite numbers.
It is obvious that  $m$ can be put equal to
$p-1$ \cite{H}.
It follows from (\ref{6}) that
\begin{equation}\label{10}
w^af^{(p)b}_{n+1}+w^bf^{(p)a}_{n+1}=\bar{D}_\lambda
j^{(p-1)\lambda ab}_{n+1}.
\end{equation}
After applying the results of Appendix 1 to equation (\ref{10}),
we have:
\begin{equation} \label{11}
  \begin{array}{l}
N^2f^{(p)a}_{n+1}=w^ay^{(p)}_{n+1}+\bar{D}_\lambda
j^{(p-1)\lambda a}_{n+1},\quad
y^{(p)}_{n+1}=\left(\frac{3}{2}N-\frac{1}{2}\gamma_cw^c
\right)\gamma_bf^{(p)b}_{n+1},
 \\
j^{(p-1)\lambda a}_{n+1}=\left(\frac{2}{3}N-
\frac{1}{6}\gamma_cw^c\right)\gamma_bj^{(p-1)\lambda ab}_{n+1}.
   \end{array}
\end{equation}
Since $ngh(f^{(p)a}_{n+1})$=1, $f^{(p)a}_{n+1}=O(C^lB^{n+1-l}),
\quad n\ge1$, the functions $f^{(p)a}_{n+1}$ contain antifields $\Phi
^*_a$, $\bar{\Phi}$, hence
$$
f^{(p)a}_{n+1}=w^a\left(\frac{1}{N^2}y^{(p)}_{n+1}\right)
+\bar{D}_\lambda
\left(\frac{1}{N^2}j^{(p-1)\lambda a}_{n+1}\right).
$$
Then, let us write $f^a_{n+1}$ as
$$
f^a_{n+1}=w^a\left(\frac{1}{N^2}y^{(p)}_{n+1}\right) +D_\lambda
\left(\frac{1}{N^2}j^{(p-1)\lambda a}_{n+1}\right)+f^{[p-1]a}_{n+1}.
$$
Obviously, $f^{[p-1]a}_{n+1}$ also obey  (\ref{6}), and
the highest summary derivatives of
$C^{\alpha a}$ and $ B^{\alpha}$ in  $f^{[p-1]a}_{n+1}$ is equal to $p-1$.
Going in the same way and taking into account that $f^a_{n+1}|_{
C^{\alpha a}=B^{\alpha}=0}=0$, we arrive at
$$
f^a_{n+1}=w^ay_{n+1}+D_\lambda
j^{\lambda a}_{n+1},\quad
F^a_{n+1}=W^a\int dx y_{n+1}\equiv
W^aY_{n+1}
$$
where $y_{n+1}$ and
$j^{\lambda a}_{n+1}$ are some functions.
In Appendix 2 we show that, under additional assumptions the functions
$y_{n+1}$  can be chosen as $Sp(2)$ and  Lorentz scalars . Taking $S_{n+1}$
=$Y_{n+1}$, we can see that the $Sp(2)$ master equation is satisfied  up to
n+1 order. Accordingly, the existence of a local solution conserving the
$Sp(2)$ symmetry is proven.

\section{The arbitrariness of solutions of the $Sp(2)$ master equation}

In this section we study the arbitrariness of solutions to the $Sp(2)$ master
equation.  Before doing that, let us introduce a set of operators which we
will call the gauge (G) transformations:
\begin{equation} \label{12}
K=\exp\left\{\frac{i\hbar}{2}:\varepsilon_{ab}
[\bar{\Delta}^b, [\bar{\Delta}^a,
F]]_+: \right\},
\end{equation}
where
$$
F=\sum_{n=0}\hbar^nF_n,\quad F_n=\int dx F_n^{\Sigma_1...\Sigma_n}
\frac{\delta}{\delta\Gamma^{\Sigma _1}}...\frac{
\delta}{\delta\Gamma^{\Sigma _n}},
$$
$F_n^{\Sigma_1...\Sigma_n}$
are functions of $\Gamma^\Sigma$
and of their finite order derivatives,
    $$ {\bar \Delta}^a = \int
dx(-1)^{\varepsilon_{A}}\frac{\delta}{\delta \Phi^A} \frac {\delta}{\delta
\Phi_{Aa}^*}+ \int dx\frac{i}{\hbar}\varepsilon^{ab}\Phi_{Ab}^*
    \frac{\delta}{\delta{\bar\Phi}_{A}},
$$
the symbol :...: means that all functionals like $\delta(0)$ must be set
equal to zero
\footnote{Correct definition of operators like
    (\ref{12}) and their properties will be given in \cite{vts}.}.

G transformations have the following properties:

1. A product of G transformations is a  G transformation.

2. Given a functional $S(\Gamma^\Sigma)$, we construct the functional
$S^\prime( \Gamma^\Sigma)$ in accordance with the rule
 \begin{equation}
\label{13}
 \exp\left[\frac{i}{\hbar}S^\prime\right]=
 K\exp\left[\frac{i}{\hbar}S\right].
\end{equation}
If $S$ is a local functional, then $S^\prime$ also is
a local functional;
if $ S $ does not depend on $\hbar$,
then $S^\prime$ has the same property;
if $ S $ satisfies the $Sp(2)$  master equation,
then $S^\prime$
satisfies the $Sp(2)$ master equation.
Two functionals  $S $ and $S^\prime $ are called gauge equivalent if they are
related by (\ref{13}). In \cite{blt} it is shown that G transformations
 do not change the physical contents of the theory.
Let us proceed by studying the general solution to
the $Sp(2)$ master equation in
first order on
 $C^{\alpha a}, B^\alpha$.
The general form of the functional, that is
 first order in
 $C^{\alpha a}, B^\alpha$,
conserves ngh and is $Sp(2)$ a scalar is
\begin{equation}\label{14}
S_1=\int dx\left(
C^{\alpha a}\Lambda^i_\alpha
\varphi^*_{ia}+
\varepsilon^{ab} B^\beta\Lambda^\alpha_\beta
C^*_{\alpha b|a}(-1)^{\varepsilon_\beta}+
B^\alpha
\bar{\Lambda}^i_\alpha
\bar{\varphi}_i
(-1)^{\varepsilon_i+\varepsilon_\alpha}+
\frac{1}{2}
\varepsilon^{ab}B^\alpha\hat{
\Lambda}^{ij}_\alpha \varphi^*_{jb}\varphi^*_{ia} \right),
\end{equation}
where $\Lambda^{i}_\alpha, \Lambda^{\beta}_\alpha,
\bar{\Lambda}^{i}_\alpha$
are LFODO's,
$$
\hat{\Lambda}^{ij}_\alpha
A_jB_i=\sum_{k, l}\lambda_\alpha^{i\mu_1...\mu_k|j\nu_1...\nu_l}
\partial_{\nu_1...\nu_l}A_j\partial_{\mu_1...\mu_k}B_i,
$$
$\lambda_\alpha^{i\mu_1...\mu_k|j\nu_1...\nu_l}$
being functions, depending on $\varphi^i$
and their finite order derivatives.
The G transformation on $S $ with
$F=-(-1)^{\varepsilon_i}\bar{\varphi_i}\bar{\varphi_j}
\hat{\Lambda}^{ij}_\alpha B^\alpha$ removes the term
$ \frac{1}{2}\varphi^*_{ia}\varphi^*_{jb}
\varepsilon^{ab}\hat{\Lambda}^{ij}_\alpha B^\alpha
$
from $S_1$
\footnote{In \cite{blt}, it was postulated that $\hat{\Lambda}^{ij}_\alpha=0$.
In fact, we see, that $\hat{\Lambda}^{ij}_\alpha$
can be removed by a G transformation.}.

The $Sp(2)$ master equation for (\ref{14}) leads to
$$
\Lambda^i_\alpha
L_i=0,\quad
\bar{\Lambda}^i_\alpha =\Lambda^\beta_\alpha \Lambda^i_\beta
$$
Due to the regularity assumptions one has:
$$
\Lambda^i_\alpha=m^\beta_\alpha R^i_\beta+\hat{M}^{ij}_\alpha L_j,\quad
\hat{M}^{ij}_\alpha A_j=\sum_{k, l}
m^{i\mu_1...\mu_k|j\nu_1...\nu_l}_\alpha
\partial_{
\nu_1...\nu_l}A_j\partial_{\mu_1...\mu_k},
$$
$$
m^{i\mu_1...\mu_k|j\nu_1...\nu_l}_\alpha=-(-1)^{\varepsilon_i\varepsilon_j}
m^{j\nu_1...\nu_l|i\mu_1...\mu_k}_\alpha,
$$
where $m^\beta_\alpha$ are LFODO's.
The expression $
C^{\alpha a}\hat{M}^{ij}_\alpha L_j
\varphi^*_{ia}
$
can be compensated by the G transformation with
$F=(1/2)C^{\alpha a}M^{ij}_\alpha\varphi^*_{ia}\bar{\varphi}_j$.
Note, that $S|_{\Phi^*_{A2}=\bar{\Phi}_A=0}$
is a solution to the master equation. We suppose that it is a proper solution,
i.e. the equations
$$
m^{T\alpha}_\beta C^{\beta a}=0,\quad \Lambda^{T\alpha}_\beta B^\beta=0
$$
have only trivial solutions, hence there exist LFODO's
$m^{-1\beta}_\alpha$, $\Lambda^{-1\beta}_\alpha$.
After the change of variables
$$
C^{\prime\alpha a}=
(-1)^{\varepsilon_\alpha+\varepsilon_\beta}
m^{T\alpha}_\beta C^{\beta a},\quad
C^{\prime*}_{\alpha b|a}=
m^{-1\beta}_\alpha
C^{*}_{\beta b|a},\quad
\bar{C}^\prime_{\alpha a}=
m^{-1\beta}_\alpha
\bar{C}_{\beta a},\quad
B^{\prime\alpha}=
m^{T\alpha}_\beta \Lambda^{T\beta}_\gamma
B^\gamma(-1)^{\varepsilon_\gamma+\varepsilon_\alpha} ,\quad
$$
$$
B^{\prime*}_{\alpha a}= \Lambda^{-1\beta}_\alpha m^{-1\gamma}_\beta
B^*_{\gamma a} ,\quad \bar{B}^\prime_\alpha= \Lambda^{-1\beta}_\alpha
m^{-1\gamma}_\beta \bar{B}_\gamma $$
(which conserves the $Sp(2)$ master equation and can be represented as a G
transformation)
we transform $S_1$ to the form (\ref{9}).

Given two solutions $S$ and $S^\prime$ to the $Sp(2)$ master equation
(\ref{1})  with the same boundary condition ${\cal S}(\varphi)$,
we suppose that they are G equivalent up to order  n.
Performing the G
transformation on
$S^\prime$, we can write
$$
S_{[n]}=S^\prime_{[n]}
,\quad
S^\prime_{n+1}=S_{n+1}+\Delta S_{n+1}
$$
where $S_1$ is given by (\ref{9}). Then $\Delta S_{n+1}$ satisfies
the equation $W^a\Delta S_{n+1}=0$ or $w^a\Delta s_{n+1}=D_\mu j^\mu_{n+1}$.
From the results of section 2 and Appendix 1, we conclude that:
$$
\Delta S_{n+1}=W^2W^1X_{n+1}.
$$
$\Delta S_{n+1}$ can be removed by applying the G
transformation to $S$ with $F=X_{n+1}$.
Applying the induction method, we conclude that the
general local solution to the $Sp(2)$ master equation, conserving
$Sp(2)$ symmetry and ngh can be represented in the form
$$
\exp\left(\frac{i}{\hbar}S\right)=\exp\left\{\frac{i\hbar}{2}:\varepsilon_{ab}
[\bar{\Delta}^b, [\bar{\Delta}^a,
X]]_+: \right\}\exp\left(\frac{i}{\hbar}S_c\right),
$$
where $S_c$ is a special solution.

{\bf Acknowledgements}

The work of S. S. S. is  supported by Russian Foundation
for Basic Researches under the Grant 96--02--17314 and by
Human Capital and Nobility Program of the European Community
under the Projects INTAS 96--0308. I.T. is partially supported
by Russian Foundation for Basic Researches under the Grant
96--01--00482 and by Human Capital and Nobility Program of the
European Community under the Project RFBR--INTAS --95829.

\section*{Appendix 1}

Here we show how equation (\ref{10}) may be solved.

Let the operators $w^a, $ $\gamma_a, $ N and $d$ define an algebra:
$$
w^aw^b+w^bw^a=0,\quad
\gamma_a\gamma_b+\gamma_b\gamma_a=0,\quad
w^a\gamma_b+\gamma_bw^a=\delta^a_bN,\quad
[w^a, d]=[\gamma_a,d]=0.
$$
Consider equation
$$
    w^af=dj^a.
    $$
By using the algebra we have
$$
    N^2f=N(w^2\gamma_1f+d\gamma_1j^2)=dN\gamma_2j^2+w^2w^1\gamma_1\gamma_2f+
    dw^2\gamma_2\gamma_1j^1=w^2w^1\gamma_1\gamma_2f+
    d\left(\frac{1}{2}(N +\gamma_aw^a)\gamma_bj^b\right),
$$
or
$$
N^2f=w^2w^1y+ dj.
$$
Now, consider the equation
\begin{equation}\label{15}
\sum_{i=1}^{l+1}w^{a_i}f^{a_1...{\not a_i}...a_{l+1}}=
d j^{
a_1...a_{l+1}}
\end{equation}
for the function $f^{a_1...a_l}$, symmetric in its indices
$a_i$.
 Multiplying (\ref{15}) by
$w^b\varepsilon_{ba_{l+1}}$
and
 $\gamma_{a_{l+1}}$, we arrive at
\begin{equation}\label{16}
(l+2)w^2w^1f^{a_1...a_l}=
d w^c\varepsilon_{cb}j^{
a_1...a_lb}
\end{equation}
and
\begin{equation}\label{17}
\gamma_bw^bf^{a_1...a_l}+lNf^{a_1...a_l}-
\sum_{i=1}^{l}w^{a_i}\left(\gamma_bf^{a_1...{\not a_i}...a_lb}\right)=
d \left(\gamma_bj^{a_1...a_lb}\right)
\end{equation}
respectively.
After multiplying (\ref{17}) by $\gamma_cw^c$
and taking into account the expressions
$$
\gamma_cw^c\gamma_bw^b=N\gamma_bw^b+2\gamma_1\gamma_2w^2w^1,\quad
\gamma_cw^cw^a=
w^a(\gamma_cw^c-N),
$$
we have
\begin{equation}\label{18}
(l+1)N\gamma_cw^cf^{a_1...a_l}=
\sum_{i=1}^{l}w^{a_i}\left[(\gamma_cw^c-N)
\gamma_bf^{a_1...{\not a_i}...a_lb}\right]+
d\left[\left(\frac{l}{l+2}\gamma_cw^c+
\frac{2}{l+2}N\right) \gamma_bj^{
a_1...a_lb}\right],
\end{equation}
where in derivation of (\ref{18}) use has been made of the equality
$\gamma_1\gamma_2w^b\varepsilon_{bc}=(\gamma_bw^b-N)\gamma_c$.
Substituting (\ref{18}) into (\ref{17}), we obtain
\begin{equation}\label{19}
l(l+1)N^2f^{a_1...a_l}=
\sum_{i=1}^{l}w^{a_i}\left[((l+2)N-\gamma_cw^c)
\gamma_bf^{a_1...{\not a_i}...a_lb}\right]+
d\left[\left(\frac{l(l+3)}{l+2}N-\frac{l}{l+2}
\gamma_cw^c\right)
\gamma_bj^{ a_1...a_lb}\right]
\end{equation} or
$$
N^2f^{a_1...a_l}=
\sum_{i=1}^{l}w^{a_i}y^{a_1...{\not a_i}...a_l}+
d j^{ a_1...a_l}.
$$
Thus, (\ref{11}) follows from (\ref{19}) with $l=1$, $d\to
\bar{D}_\mu, $ $j^a\to j^{a\mu}$.

\section*{Appendix 2}

Here we show that the functionals $Y_n$ can be chosen
$Sp(2)$ and (under additional assumptions) Lorentz scalars.
We closely follow \cite{p} in our arguments .

Let $T^\sigma$ be transformation generators of the fields $\Gamma^\Sigma$,
defining a semi-simple algebra {\bf g}.
Let the action of $T^\sigma$ on any local functional
 $A=\int dx a$ be:
\begin{equation}\label{20}
T^\sigma A=\int dx t^\sigma a,
\end{equation}
\begin{equation}\label{21}
t^\sigma
=\sum_{k=0}
t^{\sigma}_{(k)}{}^{\mu_1...\mu_k}_\Sigma{}^{\Sigma^\prime}_{\nu_1...\nu_k}
\partial_{\mu_1...\mu_k}\Gamma^\Sigma\frac{\partial}{\partial
(\partial_{\nu_1...\nu_k}\Gamma^{\Sigma ^\prime})},
\end{equation}
where
$
t^{\sigma}_{(k)}{}^{\mu_1...\mu_k}_\Sigma{}^{\Sigma^\prime}_{\nu_1...\nu_k}
$
are constant matrices, $t^\sigma$ define the same algebra {\bf g}
and
\begin{equation}\label{22}
[t^\sigma, w^a]=\tau^{\sigma a}_bw^b,\quad [t^\sigma, w^2w^1]=0,
\quad [t^\sigma, D_\mu]=a^{\sigma \nu}_\mu D_\nu,
\end{equation}
with constant matrices
$\tau^{\sigma a}_b$,  $a^{\sigma \nu}_\mu$.
Next, we consider the functionals
\begin{equation}\label{23}
F^a=W^aY,\quad Y=\int dx y,
\end{equation}
obeying the conditions
\begin{equation}\label{24}
T^\sigma F^a=\tau^{\sigma a}_bF^b.
\end{equation}
From  (\ref{24}), it follows that:
$$
w^a t^\sigma y=D_\mu j^{\mu a\sigma},
$$
with some functions
$j^{\mu a\sigma }$.
We suppose that there exists an operator $N^{-1}$ defined on the functions
$t^\sigma y$.  Taking into account the results of Appendix 1, one has:
\begin{equation}\label{25}
t^\sigma y=w^2w^1z^\sigma+ D_\mu j^{\mu\sigma}.
\end{equation}
Every subspace of the fixed power uniform polynomials in the variables
 $\Gamma^\Sigma$
and their finite order derivatives
defines a finite completely reducible representation of $t^\sigma$.
The expansion of $y$ into irreducible representations reads
$$
y=y_0+\sum_{R\neq 0}y_R,
$$
where $y_0$ belongs to the trivial representation.
Equation (\ref{25}) can be represented as follows:

\begin{equation}\label{26}
\sum_{R\neq 0}t^\sigma y_R=w^2w^1z^\sigma+ D_\mu j^{\mu\sigma}.
\end{equation}
The equation
(\ref{26}) gives

$$
y_R=w^2w^1z_R+ D_\mu j^{\mu}_R,
$$
with some functions
$z_R$ and $j^{\mu}_R$,
due to the standard arguments.
Hence, we obtain
$$
y=y_0+w^2w^1z+ D_\mu j^{\mu}
$$
with some functions
$z$ and $j^{\mu}$.
Thus, we can take
$Y_0=\int dx y_0$ as $Y$. In general case, $Y$ is invariant up to
a G transformation.
Obviously, the generators of the $Sp(2)$
transformations obey the conditions
(\ref{20}), (\ref{21}), (\ref{22}).

Now let us discuss the problem of Lorentz invariance of the functional $Y$.
The Lorentz transformation generators are:
$$
M_{\mu\nu}=\int dx\left((x_\mu\partial_\nu-x_\nu\partial_\mu)
\Gamma^\Sigma\frac{\delta}{\delta\Gamma^\Sigma}+M^{\Sigma^\prime}_
{\Sigma\mu\nu}\Gamma^\Sigma
\frac{\delta}{\delta\Gamma^{\Sigma ^\prime}}\right),
$$
where  the Lorentz transformation  matrix $M^{i^\prime}_{i\mu\nu}$
of the original fields $\varphi^i$ are known. The elements of the matrices for the
remaining fields will be defined below.
For simplicity, we consider  classical action ${\cal S}(\varphi)$
which does not depend explicitly on $x^\mu$.
Then all functions $s_k$, $f^a_n$ can be chosen independent of $x$
(see section 2). On these functions the operators $M_{\mu\nu}$
and $D_\mu$ are
$$
M_{\mu\nu}
 A=\int dx m_{\mu\nu} a,\quad
m_{\mu\nu}
=\sum_{k=0}
M_{(k)}{}^{\mu_1...\mu_k}_\Sigma{}^{\Sigma^\prime}_{\nu_1...\nu_k|\mu\nu}
\partial_{\mu_1...\mu_k}\Gamma^\Sigma\frac{\partial}{\partial
(\partial_{\nu_1...\nu_k}\Gamma^{\Sigma^\prime})},
$$
$$
M_{(k)}{}^{\mu_1...\mu_k}_\Sigma{}^{\Sigma^\prime}_{\nu_1...\nu_k|\mu\nu}
=\frac{1}{k}\sum_{i=1}^k \sum_{j=1}^k\frac{1}{k-1}
M_{(k-1)}{}^{\mu_1...{\not\mu_i}...\mu_k}_\Sigma{}^{\Sigma^\prime}_
{\nu_1...{\not\nu_j}...\nu_k|\mu\nu}
\delta_{\nu_j}^{\mu_i}
+
$$ $$
\frac{1}{k}\sum_{i=1}^k
{\tilde M}_{(k-1)}{}^{\mu_1...{\not\mu_i}...\mu_k}_\Sigma
{}^{\Sigma^\prime}_{\nu_1...\nu_{k-1}|\mu\nu}
\delta_{\nu_k}^{\mu_i}, \quad k\ge 2,
$$
$$
\lambda^{\alpha}_{\beta|\mu\nu}=\eta_{\mu\beta}\delta^\alpha_\nu-
\eta_{\nu\beta}\delta^\alpha_\mu,\quad
M_{(1)}{}^{\mu_1}_\Sigma{}^{\Sigma^\prime}_{\nu_1|\mu\nu}=
\lambda_{\nu_1}^{\mu_1}\delta_{\Sigma}^{\Sigma^\prime}+
M^{\Sigma^\prime}_{\Sigma\mu\nu}\delta_{\nu_1}^{\mu_1},\quad
m_{\mu\nu}\varphi^i=M^{i}_{i^\prime\mu\nu}\varphi^{i^\prime}.
$$
$$
D_\mu=\displaystyle\sum_{k=0}\partial_{\mu\mu_{1}...\mu_{k}}\Gamma^\Sigma
\displaystyle{\partial\over\partial(\partial_{\mu_{1}...\mu_{k}}\Gamma^\Sigma)},
$$
hence
$[m_{\mu\nu},D_\lambda]=\lambda^\sigma_{\lambda|\mu\nu}D_\sigma$.
The Lorentz invariance of the original action leads to
$m_{\mu\nu}L_i=-M^{i^\prime}_iL_{i^\prime}$.
It is natural to define
$$
m_{\mu\nu}\varphi^*_{ia}=-M^{i^\prime}_{i\mu\nu}\varphi^*_{i^\prime a},\quad
m_{\mu\nu}\bar{\varphi}_{i}=-M^{i^\prime}_{i\mu\nu}\bar{\varphi}_{i^\prime
}.
$$
Next, we suppose that the gauge
transformation generators are chosen by Lorentz covariant way ,i.e.,
$$
m_{\mu\nu}R^{i}_\alpha=M^{i}_{i^\prime\mu\nu}R^{i^\prime}_{\alpha}
-M^{\alpha^\prime}_{\alpha\mu\nu}R^{i}_{\alpha^\prime},
$$
where
$
M^{\alpha^\prime}_{\alpha\mu\nu}
$
are  generators of a finite dimensional representation of the algebra o(3, 1)
(gauge parameters transform according to this representation).
We will suppose, that the Lorentz transformation of the fields
$
C^{\alpha b}, $ $B^\alpha, $ $ C^*_{\alpha
b|a}, $ $B^*_{\alpha a}, $ $\bar{C}_{\alpha b}, $ $\bar{B}_{\alpha}
$
has the following form
$$
m_{\mu\nu}(C^{\alpha b}, B^\alpha)=
M^{\alpha}_{\alpha^\prime\mu\nu}
(C^{\alpha^\prime b}, B^{\alpha^\prime}),\quad
m_{\mu\nu}(C^*_{\alpha b|a}, B^*_{\alpha a},
\bar{C}_{\alpha b}, \bar{B}_{\alpha})=
-M^{\alpha^\prime}_{\alpha\mu\nu}
(C^*_{\alpha^\prime b|a}, B^*_{\alpha^\prime a},
\bar{C}_{\alpha^\prime b}, \bar{B}_{\alpha^\prime})
$$
(the fields
$C^{\alpha b}$
and
$B^{\alpha}$
Lorentz transform in the same way as the gauge transformation parameters)
It is easy to check, that  $m_{\mu\nu}$
(and $M_{\mu\nu}$) define the algebra o(3, 1) and that the relation
$$
[w^a, m_{\mu\nu}]=0
$$
holds.
Accordingly, the generators $M_{\mu\nu}$ and $m_{\mu\nu}$
obey the conditions
(\ref{20}), (\ref{21}), (\ref{22}).
Therefore, the results of this Appendix allow us to choose $y$
as a Lorentz scalar.

\end{document}